\documentclass[aps,prb,groupedaddress,showpacs,floats,eqsecnum]{revtex4}
\usepackage[dvips]{graphicx}
\usepackage{bm}

\begin{document}

\title{Weakly Interacting Bose Gas in the
Vicinity of the Critical Point}

\author{Nikolay Prokof'ev$^{1,2}$}
\author{Oliver Ruebenacker$^{1}$}
\author{Boris Svistunov$^{1,2}$}
\affiliation{$^{1}$Department of Physics, University of
             Massachusetts, Amherst, MA 01003, USA}
\affiliation{$^{2}$Russian Research Center ``Kurchatov Institute",
             123182 Moscow, Russia}
\date{\today}

\begin{abstract} 
We consider a three-dimensional weakly interacting Bose gas
in the fluctuation region (and its vicinity) of the
normal-superfluid phase transition point. We establish  relations
between basic thermodynamic functions: density, $n(T,\mu )$,
superfluid density $n_s(T,\mu )$, and condensate density, $n_{\rm
cnd} (T,\mu )$. Being universal for all weakly interacting
$|\psi|^4$ systems, these relations are obtained from Monte Carlo
simulations of the classical $|\psi|^4$ model on a lattice.
Comparing with the mean-field results yields a quantitative
estimate of the fluctuation region size. Away from the fluctuation
region, on the superfluid side, all the data perfectly agree with
the predictions of the quasicondensate mean field theory.---This
demonstrates that the only effect of the leading
above-the-mean-field corrections in the condensate based
treatments is to replace the condensate density with the
quasicondensate one in all local thermodynamic relations.
Surprisingly, we find that a significant fraction of the density
profile of a loosely trapped atomic gas might correspond to the
fluctuation region.

\end{abstract}

% 03.75.Hh Static properties of condensates; thermodynamical, statistical and structural properties  
% 05.30.Jp Boson systems (for Bose-Einstein condensation, see 03.75.Hh)
% 67.40.-w Boson degeneracy and superfluidity of 4He

\maketitle

%%%%%%%%%%%%%%%%%%%%%%%%%%%%%%%%%%%%%%%%%%%%
\section{Introduction}
\label{sec:introduction}
%%%%%%%%%%%%%%%%%%%%%%%%%%%%%%%%%%%%%%%%%%%%

Typically, only special properties of critical systems (such as
critical exponents and amplitude ratios) are universal (see, e.g.
Ref.~\onlinecite{book}). Weakly interacting $U(1)$ models,
including dilute Bose gases (BG), are an exception from this rule
in the sense that {\it all} their thermodynamic properties are
universal close to the critical point. This unique feature allows
complete characterization of the fluctuation region and its
vicinity   for all such models without adjustable parameters. In
the recent paper \cite{our2D} devoted to two dimensional systems
we obtained the universal thermodynamic relations for the density,
$n$, superfluid density, $n_s$, and quasicondensate density,
$n_{\rm qc}$, by performing high-precision Monte Carlo simulations
of the classical $|\psi|^4$ model on a lattice. Here we present
the results of a similar study of the three-dimensional (3D) case.

Away from the critical point, the mean-field (MF) treatment of the
weakly interacting BG \cite{bogoliubov,LP,Popov} is a reliable
theoretical scheme controlled by the small dimensionless parameter
\begin{equation}
n^{1/3} a  \ll 1 \; ,
\label{a}
\end{equation}
where $a$ is the scattering length. In the pseudo-potential
approach, one models the same scattering length $a$ by introducing
a weak short-range potential, $U_{\rm psd}$, of radius $r\ll
n^{-1/3}$ and amplitude $U=\int U_{\rm psd} ({\bf r})\,  d{\bf r}
$. The relation between $U$ and $a$ is provided by the
perturbation theory (we set $\hbar =1$)
\begin{equation}
U={4\pi a \over  m  } \left( 1+
 4\pi a \int {d{\bf p} \over (2\pi)^3 } {1\over p^2} + \dots
\right) \;.
\label{U}
\end{equation}
The divergence of the second term in the ultra-violet limit is cut
at $p \sim 1/r$; it  cancels out when final answers are expressed
in terms of the scattering length \cite{bogoliubov,LP,Popov}.
Close to the Bose condensation point, when $n^{2/3} \sim mT$, one
can write Eq.~(\ref{a}) as
\begin{equation}
mU \ll 1/\sqrt{mT} \; . \label{Usmall}
\end{equation}

In the fluctuation region,
\begin{equation}
|t|=| (T_c-T)/T_c | \sim m^{3/2}T^{1/2}U \;, \label{fregion}
\end{equation}
($T_c$ is the critical temperature),  perturbative in $U$
approaches do not work \cite{Popov,Baym99,Baym00,AM2000,us,AM}.
The physics now is determined by non-perturbative coupling between
the long-wavelength components of the order parameter field,
resulting in a crossover from MF to the generic U(1)-universality
class critical behavior on approaching the critical point.

An important observation about the fluctuation region of a weakly
interacting BG is that in the small interaction limit all
$|\psi|^4$ models---quantum or classical, continuous or
discrete---allow a universal description
\cite{Popov,Baym99,our2D}. The crucial circumstance is that
long-wavelength components of the order parameter field, $\psi
({\bf r} )$, with momenta much smaller then the thermal momentum
$k_T = \sqrt{mT}$ are classical in nature (large occupation
numbers), and only components with $k \lesssim k_c = m^2 T U \ll
k_T$ are strongly coupled. In this limit the effective Hamiltonian
is given by the classical $|\psi |^4$ model
\begin{equation}
H[\psi]= \int \left\{ {1 \over 2m}|\nabla \psi|^2 + {U \over 2}
|\psi|^4 - \tilde{\mu } |\psi|^2 \right\} \, d {\bf r} \; ,
\label{psi4}
\end{equation}
where $\tilde{\mu }$ is the renormalized chemical potential for
long-wave modes. The microscopic physics of different models is
important only at momenta $k \gg k_c$, where the system behavior
is perturbative and thus may be easily accounted for analytically.

Our approach to the problem starts from observation that $n_s$,
$n_{\rm cnd}$, and $n-n_c$, are universal functions of the shifted
chemical potential, $\mu -\mu_c$, where $\mu_c(T) , n_c(T)$ are
the (model specific) critical point parameters. To find these
functions one has to solve accurately any of the weakly
interacting $|\psi|^4$ models, and we resort to Monte Carlo (MC)
simulations of the classical $|\psi |^4$ model on a lattice using
very efficient Worm-algorithm not suffering from the critical
slowing down \cite{Worm}. In 2D, the same idea was successfully
implemented in Ref.~\onlinecite{our2D}. Before that, quantum to
classical mapping was used in Refs.~\onlinecite{us,AM,PRS} to
determine the critical point parameters (both in 2D and 3D).

This approximation-free numerical solution yields an accurate
description of the system thermodynamics. In the critical region,
characterized by known exponents of the U(1) universality class,
our goal is to find amplitudes of the power-law dependencies for
all quantities in question. By comparing to the results of the MF
theory, we establish {\it quantitative} limits on the size of the
fluctuation region. Away from the fluctuation region our accurate
data unambiguously distinguish between MF theories based on the
notion of condensate and quasicondensate---the amplitude of the
order parameter field at intermediate distances---in favor of the
latter.

In the past, most non-perturbative calculations using
renormalization group (RG) and $1/N$-expansions (for the most
recent work see, e.g.,
Refs.~\onlinecite{stoof,Baym00,AM2000,kleinert,Kastening,ledowski})
concentrated on the critical temperature shift, $\Delta T_c$, and
the scattering in the results was quite substantial. In the
absence of small parameters controlling the accuracy of the
answer, the knowledge of final results is crucial in
discriminating between competing approaches and in developing
better schemes. However $\Delta T_c$, or $\Delta n_c$, is only one
of many universal properties of weakly interacting systems. The
critical chemical potential shift as well as superfluid and
condensate density behavior in the critical regime are also
universal. A reliable theoretical approach should be able to
reproduce all of them, and our results provide corresponding
benchmarks.

The paper is organized as follows. In Sec.~\ref{sec:relations} we
use the analysis of dimensions to cast thermodynamic relations for
the weakly interacting gas in the universal form. Special
attention is paid to the ultra-violet and infra-red procedures of
the chemical potential renormalization. In Sec.~\ref{sec:MF} we
render MF theory based on the quasicondensate density. In
Sec.~\ref{sec:numerical_procedure} we introduce the classical
$|\psi |^4$ model on a lattice and the simulation algorithm.   Our
results are presented and compared to the critical and MF behavior
in Secs.~\ref{sec:comparison}.  In Sec.~\ref{sec:conclusions} we
discuss our results in the context of  experiments with ultracold
gases and make comparisons with some  analytical approaches to the
fluctuation region.

%%%%%%%%%%%%%%%%%%%%%%%%%%%%%%%%%%%%%%%%%%%%%%
\section{Universal relations for weakly
interacting $|\psi|^4$ models}
\label{sec:relations}
%%%%%%%%%%%%%%%%%%%%%%%%%%%%%%%%%%%%%%%%%%%%%%

In the $U \to 0$ limit one can present simple arguments for the
typical  energy and density scales responsible for the
non-perturbative behavior at the critical point
\cite{Baym99,us,AM}. To find the momentum separating weakly and
strongly coupled modes, $k_c$, one considers the three terms in
the Hamiltonian (\ref{psi4}) and determines when all of them are
of the same order of magnitude for modes $k\le k_c$, assuming that
modes with $k>k_c$ are already taken into account in renormalized
values of the Hamiltonian parameters. This leads to the estimates
\begin{equation}
k_c^2/m \, \sim \, | \tilde{\mu} | \, \sim \, \tilde{n} U \; ,
\label{est1}
\end{equation}
where
\begin{equation}
\tilde{n} \sim  \sum_{k<k_c} |\psi _k |^2 = \sum_{k<k_c} n_k \; ,
\label{tilde_n}
\end{equation}
is the long wavelength contribution to the total density, and
$\tilde{\mu}$ is the effective chemical potential for low-energy
modes. In 3D one has $\tilde{n} \sim k_c^3 n_{k_c}$, and since
$k_c$ is separating strongly coupled long-wave harmonics from
slightly perturbed  short-wave ones, the order-of-magnitude
estimate for $n_{k_c}$ may be obtained from the ideal system
formula:
\begin{equation}
n_{k_c} \sim \frac{T}{k_c^2/2m-\tilde{\mu}}
 \sim \frac{T}{ | \tilde{\mu} | }  \; .
\label{reltnc}
\end{equation}
Substituting this back to Eqs.~(\ref{est1})-(\ref{tilde_n}) yields
\begin{equation}
k_c = m^2 T U \; ,
\; \; \; \tilde{n} \sim m^3 T^2 U  \;,
\; \; \; | \tilde{\mu} | \sim m^3 T^2 U^2
 .
\label{est2}
\end{equation}

%%%%%%%%%%%%%%%%%%%%%%%%%%%%%%%%%%%%%%%%%%%%%%%%%%%%%%%%%%%
\begin{figure}[tbp]
\includegraphics[bb=120 150 550 800, angle=-90, width=0.4\columnwidth]{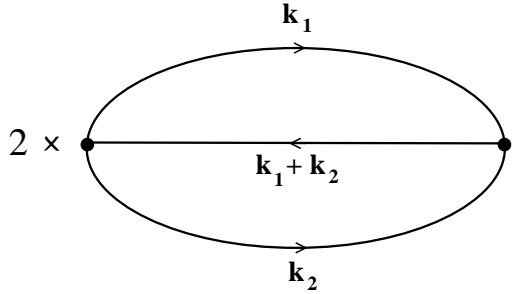}
\caption{The second order diagram for the self-energy
$\Sigma (\omega =0, p=0 )$.}
\label{fig1}
\end{figure}
%%%%%%%%%%%%%%%%%%%%%%%%%%%%%%%%%%%%%%%%%%%%%%%%%%%%%%%%%%%

The critical values of the density, or chemical potential, are, of
course, model specific. [We find it convenient to work in the
grand canonical ensemble and keep temperature fixed; in
Sec.~\ref{sec:comparison} we explain how all the results are
easily converted into more familiar temperature plots.] In fact,
in 3D the major contribution to $n_c$, and $\mu_c$, is coming from
high momenta $k \gg k_c$. However, this model specific
contribution to density does not depend on interactions in linear
order in U, and thus can be easily calculated analytically. Thus
if we count density and chemical potential from their critical
values, we expect the dependence of $n-n_c$ on $\mu-\mu_c$ to be
universal because of its long-wavelength nature. In view of
Eq.~(\ref{est2}) we write the equation of state in the
dimensionless form as
\begin{equation}
n-n_c \, =\, (m^3T^2U)\,  \lambda(X) \; , \label{lambda}
\end{equation}
where $X$ is the universal control parameter
\begin{equation}
 X\, =\, {\mu -\mu_c \over m^3T^2U^2} \; ,
 \label{X}
\end{equation}
with the typical variation across the fluctuation region of order
unity. [For the gas in a slowly varying external potential,
$V({\bf r})$, the function $\lambda(X)$ describes the spatial
density profile, provided $\mu$ in Eq.~(\ref{X}) is replaced with
$\mu - V({\bf r})$.] Similarly,
\begin{equation}
n_{\rm cnd} \, =\,  (m^3T^2U)\,  f_0(X)  \; , \label{f0}
\end{equation}
\begin{equation}
n_s \, =\,  (m^3T^2U) \, f_s(X) \; , \label{fs}
\end{equation}
Establishing universal---for all weakly-interacting $|\psi|^4$
models---functions $\lambda$, $f_0$, and $f_s$, and explaining how
they work for the quantum Bose gas is the main goal of this paper.

First, we note that $n_c$ and $\mu_c$ for the Bose gas
have been already determined in previous MC studies \cite{us,AM,AM2002}.
The interaction induced shift of the critical density is universal
\begin{equation}
n_c = n_c^{(0)} - C  m^3T^2U \; ,
\;\;\; C=0.0142(4) \;,
\label{nc}
\end{equation}
where $n_c^{(0)}$ is the critical density of the non-interacting
system; $n_c^{(0)}=(mT/3.313)^{3/2}$ for the Bose gas. [We quote
the mean of the results presented in Refs.~\cite{us} and \cite{AM}
which overlap within the error bars.]

Before quoting the result for $\mu_c$, we would like to discuss
its subtle structure  \cite{AM,Baym2001,AM2002}. Linear in $U$
terms are accounted for by the MF expression $2nU$. There are two
distinctive contributions to $\mu_c$ which are quadratic in $U$.
One is universal and comes from strongly coupled critical
fluctuations. The other one is perturbative and comes from the
second-order diagram (see Fig.~\ref{fig1}) for the self-energy,
$\Sigma^{(2)} ( \omega =0, p=0 )$:
\begin{eqnarray}
 \Sigma^{(2)} &=& -2U^2T^2 \int \int{ d{\bf k}_1 d{\bf k}_2 \over (2\pi )^6 }
 \sum_{s_1 s_2} {1 \over
 [i\omega_{s_1}-\epsilon_{k_1}]
 [i\omega_{s_2}-\epsilon_{k_2}]
 [i(\omega_{s_1}+\omega_{s_2})-\epsilon_{{\bf k}_1+{\bf k}_2}  ] }
 \nonumber \\
 &=& {2U^2 \over (2\pi )^6}  \int \int
 { d{\bf k}_1 d{\bf k}_2 \over
 \epsilon_{k_1}+\epsilon_{k_2}-\epsilon_{{\bf k}_1+{\bf k}_2}  }
 \left[ N_{k_1}N_{k_2}-N_{{\bf k}_1+{\bf k}_2}
 \left( N_{k_1}+ N_{k_2}+1 \right)  \right]
  \;  \;\;\;\mbox{(quantum)} \;,
\label{SigmaQ}
\end{eqnarray}
where $\epsilon_{{\bf k}}$ is the dispersion law, $\omega_{s}=2\pi T s$
is the Matsubara frequency, and
$N_{k} = [e^{\epsilon_{k}/T } -1 ]^{-1}$ is the Bose distribution function.
The corresponding classical expression is obtained by considering only the
zero-frequency term ($s_1=s_2=0$):
\begin{equation}
 \Sigma^{(2)} =  {2U^2T^2 \over (2\pi )^6}  \int \int
 { d{\bf k}_1 d{\bf k}_2 \over
 \epsilon_{k_1} \epsilon_{k_2} \epsilon_{{\bf k}_1+{\bf k}_2}  }  \;
 \;\;\;\mbox{(classical)} \;,
\label{SigmaC}
\end{equation}

The self-energy integrals are divergent.
In the infrared limit the divergence is
logarithmic
\begin{equation}
 \stackrel{k \ll k_T} \longrightarrow \;
      {m^3T^2U^2\over \pi^2 } \int dk/k  \;.
 \label{infrared}
\end{equation}
For the Bose gas ($\epsilon_k=k^2/2m $) it is also divergent in the
ultra-violet limit as a power law
\begin{equation}
 \stackrel{k \gg k_T}\longrightarrow \;
  -{2U^2 \over (2\pi )^6}  \int \int
 { d{\bf k}_1 d{\bf k}_2 N_{{\bf k}_1+{\bf k}_2} \over
 \epsilon_{k_1}+\epsilon_{k_2}-\epsilon_{{\bf k}_1+{\bf k}_2}  }
 \approx  -{2U^2m \over (2\pi )^6} \int d{\bf k} N_{k}
                                   \int  {d{\bf p} \over p^2 }
  = -2nU  \left( 4\pi a \int {d{\bf p} \over (2\pi)^3 } {1\over p^2} \right)
   \equiv \Sigma^{(2)}_{UV} \;.
\label{ultraviolet}
\end{equation}
We cast the last expression in the form which immediately relates
it to the definition of the pseudo-potential, Eq.~(\ref{U}), i.e.
in the sum $2nU + \Sigma^{(2)}$ the ultra-violet divergences
cancel each other for terms $\propto U^2$. Identically, the same
result is obtained if in all final answers we simply substitute $U
\to 4\pi a/m$ and use
\begin{equation}
\tilde{\Sigma }^{(2)}=\Sigma^{(2)}-\Sigma^{(2)}_{UV}
~~~~~~\mbox{(quantum)} \; , \label{SUV}
\end{equation}
as the second-order self-energy for the quantum Bose gas. In what
follows we employ this well established trick
\cite{bogoliubov,LP,Popov}. Obviously, in classical models quantum
corrections to $U$ are absent. Correspondingly, there are no
divergencies in $\Sigma^{(2)}$ apart from the logarithmic one, and
the original expression for the self-energy should be used:
$\tilde{\Sigma }^{(2)}=\Sigma^{(2)}$.

The logarithmic divergence (\ref{infrared}) can be simply truncated
at some infra-red energy scale, or regularized \cite{AM,AM2002}.
Whatever the procedure, as long as it is the same for all weakly-interacting models,
the difference between the true value
of $\mu_c$ and $2n_cU+\tilde{\Sigma }^{(2)}$ is coming from
long-wave modes and thus is universal, while the $\tilde{\Sigma }^{(2)}$
term is model specific.
In this paper we introduce an
explicit infra-red cutoff by adding a constant
$\kappa = k_c^2/2m = m^3T^2U^2/2$ to the dispersion law, i.e.
$\epsilon_k \to \epsilon_k+\kappa$ everywhere in Eqs.~(\ref{SigmaQ})
and (\ref{SigmaC}). Then
\begin{equation}
\tilde{\Sigma }^{(2)}\, =\, - \,{m^3T^2U^2 \over \pi^2} \,  \ln
\left(5.9030\,   \sqrt{m^3 T U^2} \right) \, =\,  {m^3T^2U^2 \over
\pi^2} \,  \ln \left( 5.9030 \, { k_T \over k_c } \right) \;,
\label{SigmaQn}
\end{equation}

The numerical value of $\mu_c$ has been determined in
Refs.~\onlinecite{AM,AM2002}. For the quantum gas it reads
\begin{equation}
\mu_c=2n_c^{(0)}U + {m^3T^2U^2 \over \pi^2} \ln \left( 0.4213(6)
\, { k_T \over k_c}  \right) \equiv 2n_c U + \tilde{\Sigma }^{(2)}
- (m^3 T^2 U^2)\, \theta_0 \; , \label{muc}
\end{equation}
where
\begin{equation}
\theta_0=(1/\pi^2) \ln (5.9030/0.4213)-2C \, = \, 0.239(1) \; .
\label{theta0}
\end{equation}

In the MF theory, the definition of the
effective chemical potential $\tilde{ \mu }$
involves subtraction of the self-energy contributions
\begin{equation}
\tilde{ \mu }=\mu-2nU-\tilde{\Sigma }^{(2)} \;.
\label{tmu}
\end{equation}
This quantity is model-independent, and we find it
convenient to introduce the corresponding universal function $\theta (X)$
\begin{equation}
2nU+\tilde{\Sigma }^{(2)}-\mu =m^3T^2U^2 \theta(X)  \;.
\label{theta}
\end{equation}
It is, of course, directly related to the equation of state
function $\lambda(X)$
\begin{equation}
\theta(X)=  2\lambda(X)-X + \theta_0 \; . \label{theta-lambda}
\end{equation}

If we were to change the infra-red cutoff in the definition of
self-energy, $\kappa \to \kappa'$ we would have to modify the
value of $\theta_0$ accordingly $\theta_0 \to \theta_0 +(1/2\pi^2)
\ln (\kappa /\kappa' )$.

%%%%%%%%%%%%%%%%%%%%%%%%%%%%%%%%%%%%%%%%%%%%%%
\section{ Mean-field Theory}
\label{sec:MF}
%%%%%%%%%%%%%%%%%%%%%%%%%%%%%%%%%%%%%%%%%%%%%%
Away from the fluctuation region our simulation is supposed to
reproduce the known perturbative results for a weakly interacting
Bose gas. That is we need  $\theta$, $\lambda$, $f_0$, and $f_s$
as functions of $X$, the dimensionless variable $1/X$ playing the
role of the small parameter that guarantees the applicability of
perturbative treatment. The leading terms in the perturbative
expansion away from the fluctuation region are $\sim |X|$, the
next-order terms are $\sim \sqrt{|X|}$. We confine ourselves to
considering only these terms, ignoring the higher order
corrections (in fact, we are not aware of existing results for
them). Hence, our numerics is expected to agree with the known
analytic results only at large enough $|X|$, and only up to some
constants.

This degree of accuracy can be achieved within the MF treatment
(plus an extra care of the higher-energy logarithmic correction to
the chemical potential) both on the normal and superfluid sides.
In the superfluid region, however, it is essential, that MF be
based on the {\it quasicondensate}, rather than genuine
condensate. In the condensate-based treatments the same accuracy
is achieved only if beyond-the-MF corrections are taken into
account. The logarithmic correction to the chemical potential,
$\tilde{\Sigma }^{(2)}$, has been already discussed in
Sec.~\ref{sec:relations}. For practical purposes, this correction
is not large---unless the gas parameter $n^{1/3}a$ is
exponentially small, and taking it into account really makes sense
only if all the $X$-independent terms are accounted for as well.
Theoretically, however, this correction involves the ultraviolet
physics and thus is model-specific. Hence, if we ignore it, then
we cannot render our answers model-independent. On the good side,
we do not need to calculate $\tilde{\Sigma }^{(2)}$ more
accurately than it was done in Sec.~\ref{sec:relations}, since
that would lead to the higher-order corrections which we ignore
here. Thus, we simply add the term (\ref{SigmaQn}) to MF
expression  for the chemical potential of the Bose gas.

\subsection{Asymptotic behavior in the normal phase ($X \to - \infty$) }
%%%%%%%%%%%%%%%%%%%%%%%%%%%%%%%%%%%%%%%%%%%%%%%%%%%%%%%%%

In the normal region $X<0$ we have $f_0$ and $f_s$ identically
zero. Hence, the only quantity we have to look at is the equation
of state $\lambda (X)$ or $\theta(X)$. Using MF equation for the
effective chemical potential $\tilde{\mu } = \mu - 2nU-
\tilde{\Sigma }^{(2)}= -\theta m^3T^2U^2$, we calculate the
difference between the density of the interacting gas
$n(T,\tilde{\mu })$ and the critical density of the ideal gas
$n_c^{(0)}$ keeping only the leading linear in $U$ terms:
\begin{equation}
n_c^{ (0)} - n \approx \int \! \!
\frac{d{\bf k}}{(2\pi )^3}
\left[  {T \over \epsilon } -
        { T \over \epsilon + \theta m^3T^2U^2 } \right] =
{ \sqrt{\theta} \,  m^3T^2U  \over \sqrt{2} \pi } \; .
\label{dif1}
\end{equation}
We now notice that $n_c^{\rm (0)} - n = [C-\lambda(X)]m^3T^2U $,
see Eqs.~(\ref{lambda}) and  (\ref{nc}), and use
Eq.~(\ref{theta-lambda}) to complete the derivation
\begin{equation}
\theta + { \sqrt{2} \over \pi} \, \sqrt{\theta}
 = 2C+\theta (0) -X  ~~~{\rm
at}~~X \to -\infty \; .
 \label{lim00}
 \end{equation}

%%%%%%%%%%%%%%%%%%%%%%%%%%%%%%%%%%%%%%%%%%%%%%%%%%%%%%%%%
\subsection{Mean-field description of the
            superfluid region ($X \to \infty$).
            Quasicondensate}
%%%%%%%%%%%%%%%%%%%%%%%%%%%%%%%%%%%%%%%%%%%%%%%%%%%%%%%%%

The standard MF approach to the weakly interacting Bose gas deals
with the condensate density, $n_{\rm cnd}$, defined through the
one-particle density matrix,
\begin{equation}
\rho(r)= \langle \psi^{\dagger}({\bf r}) \psi({\bf 0})
\rangle
\label{rho}
\end{equation}
[$\psi$ is either the field operator (in the quantum system) or a
complex valued field (in the classical system); in the latter
case $\psi^{\dagger} \equiv \psi^*$], as
\begin{equation}
n_{\rm cnd} = \lim_{r \to \infty} \rho(r) \; . \label{cond}
\end{equation}

Well inside the region of applicability of the MF description, but
not too far from the fluctuation region, the MF theory based on
$n_{\rm cnd}$ is less accurate than the theory dealing with the
notion of {\it quasi-}condensate. Of course one can go beyond the
MF description in the condensate based theory and evaluate the
corresponding corrections. It is important, however, that the
quasicondensate MF description automatically captures the first
order corrections to the condensate MF. [In 2D the quasicondensate
MF has demonstrated perfect agreement with MC simulations away
from the fluctuation region\cite{our2D}.] We thus find it
instructive to resort here to the quasicondensate MF description.

The notion of quasicondensate was introduced by Popov\cite{Popov}
(``bare condensate" in the original version). Basically, it is
used to describe the order parameter with fluctuating phase (see
also \cite{KSS}) and implies the possibility of parameterizing the
field $\psi ({\bf r})$ for the weakly interacting system as
\begin{eqnarray}
\psi ({\bf r})  & = & \psi_0 ({\bf r}) + \psi_1 ({\bf r}) \; ,
\label{psi} \\ \psi_0 ({\bf r})& \approx &
 \sqrt{n_{\rm qc}} \; e^{i \Phi ({\bf r})} \; ,
\label{psi_0}
\end{eqnarray}
where $n_{\rm qc}$ is the quasicondensate density, and $\psi_1$ is
the Gaussian field independent of $\psi_0$. The Gaussian field
$\psi_1$ is primarily responsible for the decay of $\rho (r)$ at
distances comparable to the thermal wavelength. The relation
between $n_{\rm cnd}$ and $n_{\rm qc}$ immediately follows from
(\ref{psi})-(\ref{psi_0}):
\begin{equation}
n_{\rm cnd} = n_{\rm qc} \lim_{r \to \infty} \langle e^{i \Phi
({\bf r}) - i\Phi ({\bf 0})}\rangle \; . \label{rel_cnd}
\end{equation}
When fluctuations of the phase field, $\Phi ({\bf r})$, become
noticeable---this is precisely the case in the vicinity of the
critical point--- the difference between $n_{\rm cnd}$ and $n_{\rm
qc}$ should be taken into consideration as well.

There are strong arguments that it is $n_{\rm qc}$, and not
$n_{\rm cnd}$, that is relevant to all the characteristics of the
system, with $n_{\rm cnd}$ being just one of them. Indeed, the
physics at large distances, or small, but finite momenta,
including the spectrum of elementary excitations, is determined by
what is happening on smaller lengthscales, in other words,
``high-energy" physics ultimately determines what happens at lower
energies. The condensate is the macroscopic characteristic of the
system; it should be derived from other finite-$k$ properties, and
$n_{\rm qc}$ governs them.

This fact has been rigorously shown by one of us \cite{Svist}.
Though the analysis of Ref.~\onlinecite{Svist} has been done for
2D systems (where the notion of quasicondensate is of crucial
importance and cannot be avoided), it is applicable to the 3D case
as well. The actual treatment closely follows Popov's theory
\cite{Popov}, with the only (but important) exception that $n_{\rm
qc}$ is understood and treated as a physical quantity. [Popov
treated bare condensate as an auxiliary mathematical quantity
explicitly dependent on the momentum $k'$ separating `slow'
harmonics from the `fast' ones. Correspondingly, he attempted to
exclude this quantity from final answers to render them
$k'$-independent.---At low enough temperatures this can be easily
done by replacing $n_{qc} \to \mu / U$. However, in the high
temperature region the quantity $n_{\rm qc}(T)$ is a meaningful
physical parameter \cite{KSS}.]

One may consider Eq.~(\ref{psi_0}) as the definition $n_{\rm
qc}$---it is the modulus of the order parameter field at large
distances. This quantity appears in all MF equations just like
$n_{\rm cnd}$ does at low temperature. Then, $n_{\rm qc}$ and $T$
can be chosen as convenient independent thermodynamic parameters
specifying the state of the system, the rest of the
characteristics being expressed as functions of  $(n_{\rm qc},
T)$. The basic results are as follows \cite{KSS,Svist}:
\begin{equation}
n=n_{\rm qc} + n' \; , \label{n'}
\end{equation}
with the non-quasicondensate part of the particle density, $n'$,
given by the integral
\begin{equation}
n'=\int \frac{{\rm d}^d k}{(2\pi )^d}
 \left[ \frac{\epsilon (k) + n_{\rm qc} U -E(k)}{2E(k)}
+ \frac{\epsilon (k) \, N_E}{E(k)} \right] \; , \label{rel7}
\end{equation}
where $\epsilon(k)=k^2/2m$ is the free-particle dispersion law,
and
\begin{equation}
E(k)= \sqrt{\epsilon (k)[\epsilon (k) + 2n_{\rm qc} U]} \label{E}
\end{equation}
is the Bogoliubov quasiparticle spectrum.

For the one-particle density matrix one obtains \cite{Svist} (see
also Ref.~\onlinecite{Krasavin} for a numeric check in 2D)
\begin{equation}
\rho (r)   =   e^{-\Lambda (r)} \, \tilde{\rho}(r)  \; ,
\label{rel1}
\end{equation}
\begin{equation}
\Lambda (r)   =   \int \! \frac{{\rm d}^d k}{(2\pi )^d} \left[
1-\cos ({\bf k} \! \cdot \! {\bf r}) \right] \frac{U \, N_E}{E}
 \; ,
\label{rel2}
\end{equation}
\begin{equation}
 \tilde{\rho} (r)   =   n_{\rm qc}  +  \int \!
\! \frac{{\rm d}^d k}{(2\pi )^d} \, \cos ({\bf k} \! \cdot \! {\bf
r}) \!  \left[ \frac{\epsilon  \! + \!  n_{\rm qc} U\!  - \! E}{2E} +
\frac{\epsilon \, N_E}{E} \right] .
\label{rel3}
\end{equation}
[In Eq.~(\ref{rel3}), as well as in Eq.~(\ref{rel7}), the first
term in square brackets is $\sim U^{3/2}$ after integration, and
may be omitted in the region $T \gg nU$ addressed in the present
paper.]

That $n_{\rm qc}$ is the relevant quantity behind the long-wave
physics is clear from the structure of
Eqs.~(\ref{rel1})-(\ref{rel3}). The second term in
Eq.~(\ref{rel3}) decays first as a power law $\sim (k_Tr)^{-1}$,
and then exponentially fast, so that on large length-scales the
amplitude of the order parameter is given by $n_{\rm qc}$. Phase
fluctuations are described by $\Lambda(r)$. They are negligible at
short distances of order  $1/k_T$, but at distances  $\gg
1/\sqrt{mn_{\rm qc}U}$ their contribution to the density matrix
results in the difference between $n_{\rm qc}$ and $n_{\rm cnd}$,
see Eq.~(\ref{rel_cnd}):
\begin{equation}
n_{\rm cnd} \,  = \,  e^{-\Lambda (\infty)}\, n_{\rm qc}\; .
\label{cond_n0}
\end{equation}

The trick to solve MF equations in the vicinity of the critical
point ($1\ll X \ll k_T/k_c$) is to calculate {\it differences}
between the corresponding densities (this makes all integrals to
converge at energies $\ll T$) keeping only the universal leading
low-momenta terms. To get $n'$ we add and subtract the ideal gas
critical density
\begin{equation}
n_c^{(0)} - n' \approx \int \! \! \frac{{\rm d}^3 k}{(2\pi )^3}
\left[  {T \over \epsilon} - {\epsilon T \over E^2}  \right] =
{m^3T^2U \sqrt{f_{\rm qc}} \over \pi} \; ,\label{dn'}
\end{equation}
where
\begin{equation}
 f_{\rm qc}={n_{\rm qc} \over m^3T^2U} \, \gg \, 1  ~~~~~({\rm at}~~X \gg 1) \;
\label{lim01}
\end{equation}
is the dimensionless parameter of asymptotic expansion for all
thermodynamic quantities away from the fluctuation region.

Combining (\ref{dn'}) with (\ref{n'}) yields
\begin{equation}
 n =n_c^{(0)}+m^3T^2U \, [f_{\rm qc} - \sqrt{f_{\rm qc}}/\pi ]
    \; ,
\label{lim1}
\end{equation}
and with (\ref{nc}) we get
\begin{equation}
\lambda =f_{\rm qc} - \sqrt{f_{\rm qc}}/\pi +C  ~~~{\rm at}~~X \to
\infty \; .\label{lim1a}
\end{equation}

The relation for the condensate density immediately follows from
the asymptotic value of the phase correlator
\begin{equation}
\Lambda (\infty) \approx  \int \! \! \frac{{\rm d}^3 k}{(2\pi )^3}
\, \frac{UT}{E^2(k)} = \frac{1}{2\pi \sqrt{f_{\rm qc}} } \;
.\label{L_inf}
\end{equation}
Hence
\begin{equation}
f_0 =  f_{\rm qc}\,  e^{-1 / (2\pi \sqrt{f_{\rm qc}}) } ~~~{\rm
at}~~X \to \infty \; . \label{lim2}
 \end{equation}

To find $f_s$, we consider the standard expression for the normal
component density  \cite{LP}
\begin{equation}
 n_n = -{1 \over 3m} \int \frac{{\rm d}^3 k}{(2\pi )^3} \,
 \left[ {{\rm d  N_E} \over {\rm d E}} \right] \, k^2 \; .
\label{normal}
\end{equation}
Integrating by parts, we rewrite Eq.~(\ref{normal}) as
\begin{equation}
n_n=\frac{2(2m)^{3/2}}{3(2\pi)^2} \int_0^{\infty} {\rm d}\epsilon
\, N_E \, \frac{\partial}{\partial \epsilon}  \; \epsilon^{3/2}
\frac{\sqrt{\epsilon (\epsilon + 2n_{\rm qc} U)}}{\epsilon + n_{\rm qc}U} \; .
\label{normal2}
\end{equation}
We now subtract from $n_n$ the non-condensate density to evaluate
the integral explicitly, and use identity $n_n-n'\equiv n_{\rm qc}
- n_s$ to get
\begin{equation}
f_s =  f_{\rm qc} -  \sqrt{  f_{\rm qc}}/3 \pi  ~~~{\rm at}~~X \to
\infty \; . \label{lim3}
 \end{equation}

Finally, the quasicondensate density can be related to the
chemical potential as \cite{KSS}
\begin{equation}
 \mu = (n_{\rm qc}+2n')U + \tilde {\Sigma}^{(2)}
     =2nU - n_{\rm qc}U + \tilde {\Sigma}^{(2)} \; ,
\label{rel0}
\end{equation}
where we have added  the term $\tilde {\Sigma}^{(2)}$, in
accordance with the previous discussion. In contrast to the
analogous relation in terms of the condensate and non-condensate
densities\cite{HP}, this relation does not imply corrections of
the order $\sqrt{X}$.

Comparing Eq.~(\ref{rel0}) with the definition of
$\theta$-function,  Eq.~(\ref{theta}), we see that
\begin{equation}
\theta = f_{\rm qc}  ~~~{\rm at}~~X \to \infty \;
,\label{qc_theta}
\end{equation}
and with the $ \theta$--$\lambda$ relation (\ref{theta-lambda})
and Eq.~(\ref{lim1a}) for $\lambda$ we get a self-consistent
equation for $f_{\rm qc}$
\begin{equation}
f_{\rm qc} - 2 \sqrt{ f_{\rm qc}}/\pi \, = \,  X - \theta (0) -2C
~~~{\rm at}~~X \to \infty \; . \label{lim0}
\end{equation}
Equation (\ref{lim0}) along with Eqs.~(\ref{qc_theta}),
(\ref{lim1a}), (\ref{lim2}), and (\ref{lim3}) define $\theta$,
$\lambda$, $f_0$, and $f_s$ as functions of $X$.

We are in a position now to demonstrate that quasicondensate MF
reproduces results of the condensate-based diagrammatic technique
which accounts for leading, $\sim \sqrt{X}$, corrections to the
condensate MF answers \cite{Popov}. Indeed, in the limit $X \to
\infty $, Eqs.~(\ref{lim0}) and (\ref{lim2}) define the effective
chemical potential dependence on the condensate density as
\begin{equation}
X \approx f_0(1+\sqrt{f_0}/2\pi ) -2\sqrt{f_0}/\pi =
f_0-3\sqrt{f_0}/2\pi \; .
\end{equation}
Similarly,
\begin{equation}
f_s \approx f_0(1+\sqrt{f_0}/2\pi ) -\sqrt{f_0}/3\pi =
f_0+\sqrt{f_0}/6\pi  \; ,
\end{equation}
in complete agreement with Ref.~\onlinecite{Popov}.

%%%%%%%%%%%%%%%%%%%%%%%%%%%%%%%%%%%%%%%%%%%%%%%%%%%%%%%%%%%
\begin{figure}[tbh]
\includegraphics[bb=0 190 650 840, angle=-90, width=0.4\columnwidth]{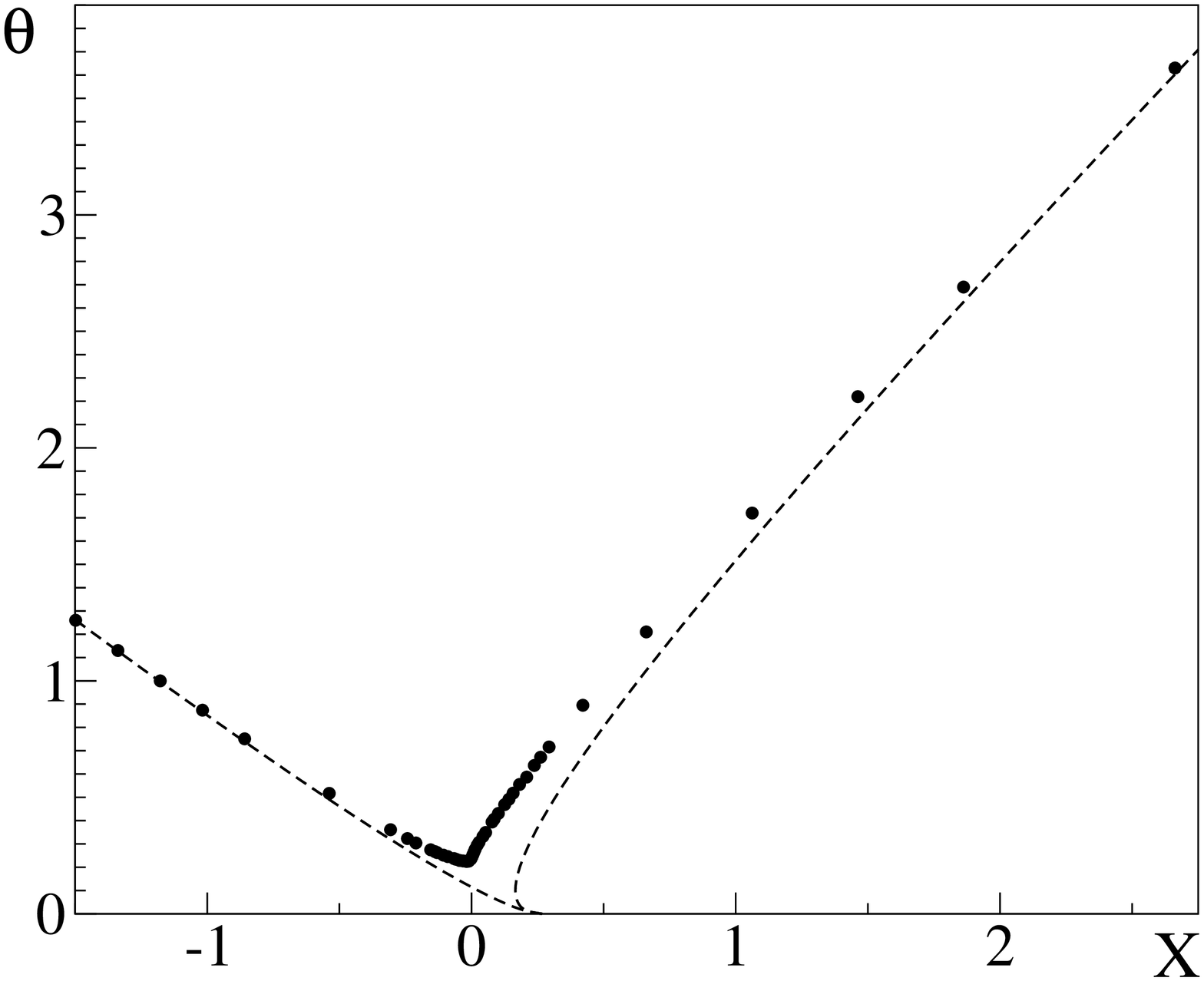}
\includegraphics[bb=0 -0 650 650, angle=-90, width=0.4\columnwidth]{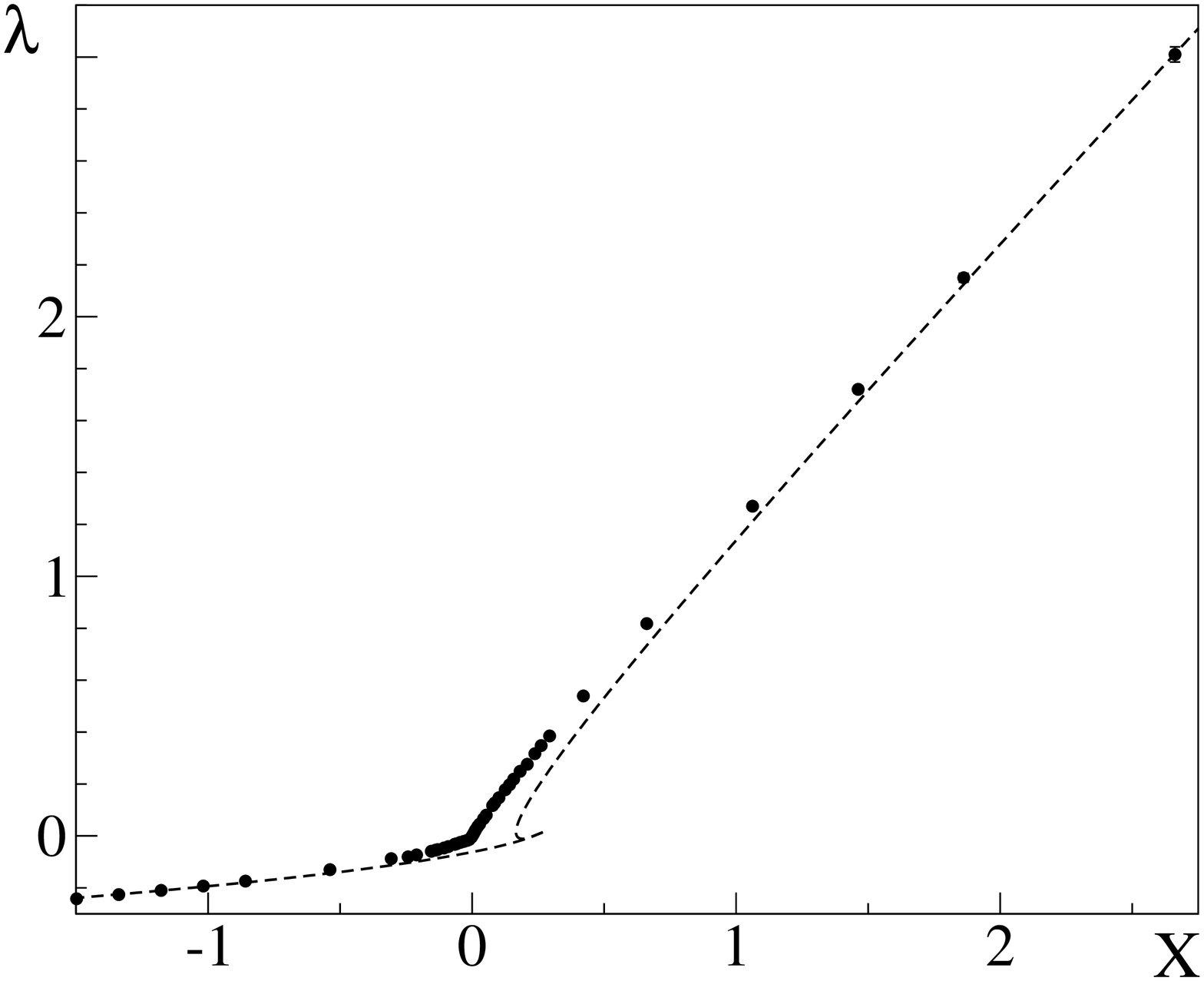}
\caption{The $\theta (X)$-function  and the
isothermal density profile $\lambda (X)$ along with the
corresponding MF expressions, Eqs.~(\ref{lim00})  and (\ref{lim0}),
shown by dashed lines.
 }
\label{figT}
\end{figure}
%%%%%%%%%%%%%%%%%%%%%%%%%%%%%%%%%%%%%%%%%%%%%%%%%%%%%%%%%%%

%%%%%%%%%%%%%%%%%%%%%%%%%%%%%%%%%%%%%%%%%%%%%%%%%%%%%%%%%%%
\begin{figure}[tbh]
\includegraphics[bb=0 190 650 840,  angle=-90, width=0.4\columnwidth]{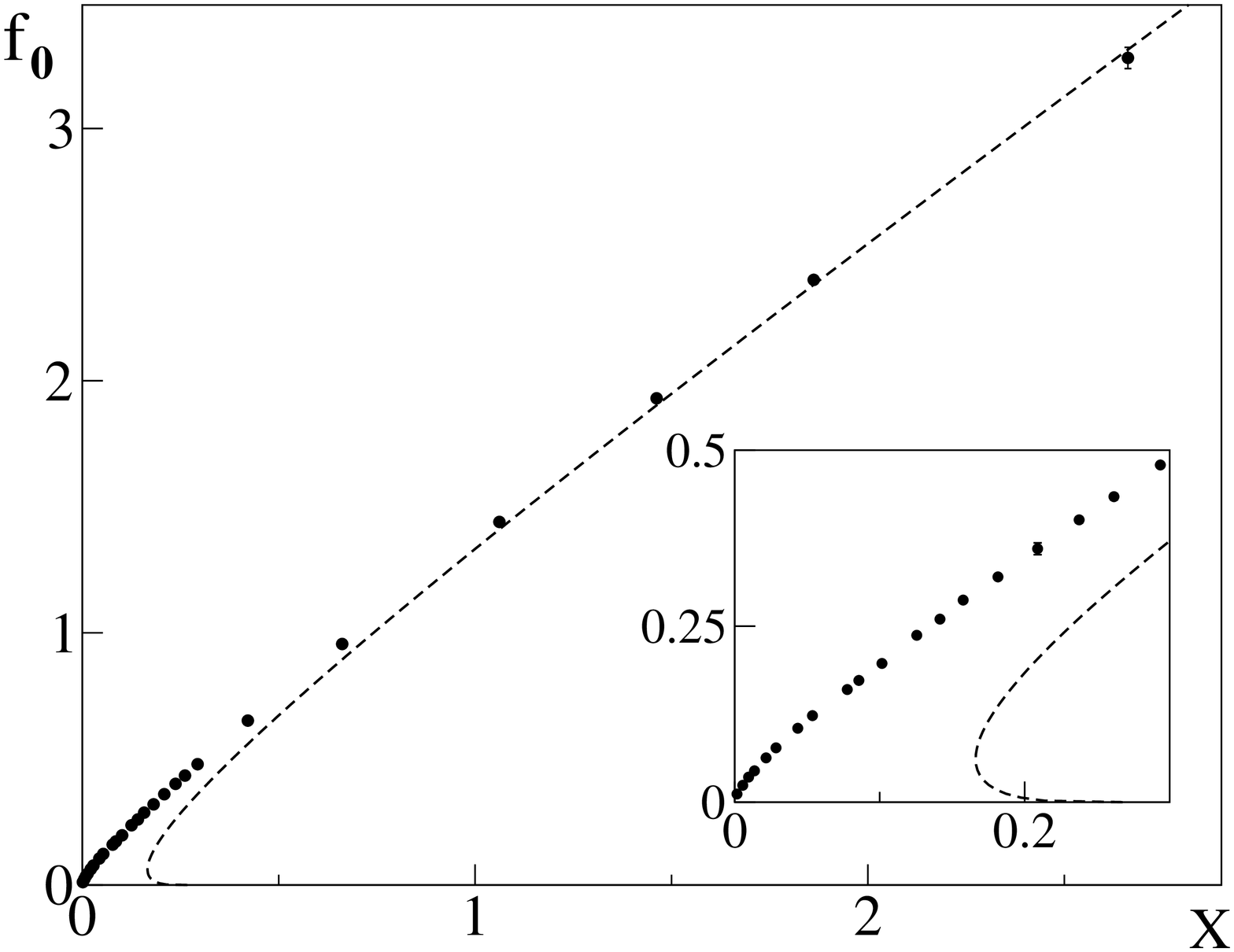}
\includegraphics[bb=0 -0 650 650 , angle=-90, width=0.4\columnwidth]{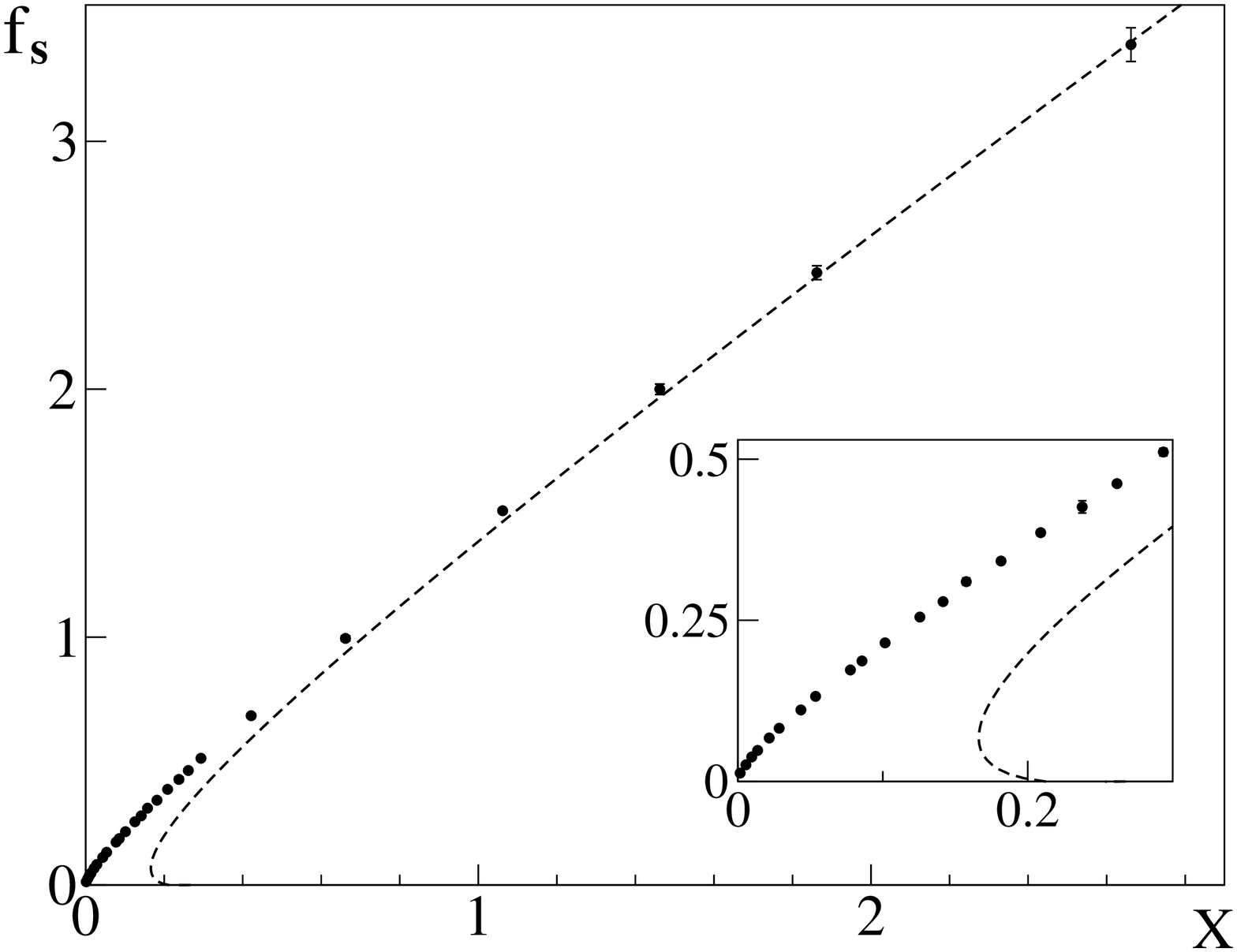}
%\includegraphics[bb= .   0    .    650, width=0.5\columnwidth]{n_cnd.ps}
%                     top    bottom
\caption{The condensate and superfluid density dependencies on $X$
and the corresponding MF predictions based on Eqs.~(\ref{lim2})
and (\ref{lim3}) shown by dashed lines. The initial parts of the
plots are shown in the insets. } \label{figN0}
\end{figure}
%%%%%%%%%%%%%%%%%%%%%%%%%%%%%%%%%%%%%%%%%%%%%%%%%%%%%%%%%%%

%%%%%%%%%%%%%%%%%%%%%%%%%%%%%%%%%%%%%%%%
\section{Lattice $|\psi|^4$ model and simulation algorithm}
\label{sec:numerical_procedure}
%%%%%%%%%%%%%%%%%%%%%%%%%%%%%%%%%%%%%%%%

As explained in Secs.~\ref{sec:introduction} and
\ref{sec:relations}, one can introduce universal functions for any
model described in the long-wave limit by the Hamiltonian
(\ref{psi4}) with small $U$. Classical lattice models are easier
to deal with numerically, and there are very efficient classical
algorithms which allow simulations of very large system sizes. In
the present study, we performed simulations of systems with up to
$128^3$ lattice points. In addition, results obtained for a
classical model directly test the idea of universality, since they
have to agree with all MF predictions formally derived for the
quantum Bose gas.

Our simulations were done for the simple cubic lattice
Hamiltonian
\begin{equation}
H = \sum_{{\bf k} \in BZ} [E({\bf k}) - \mu ] |\psi_k |^2 +
 {U\over 2 } \sum_i |\psi_i|^4\; ,
\label{H_lat}
\end{equation}
where $\psi_k$ is the Fourier transform of the complex lattice field
$\psi_i$, and
\begin{equation}
\epsilon_{\bf k} = {1 \over ma^2 } \sum_{\alpha = x,y,z } [ 2-
\cos (k_\alpha a)] \;, \label{tight}
\end{equation}
is the tight-binding dispersion law with momentum ${\bf k}$
confined within the first Brillouin zone (BZ). The self-energy
expression (\ref{SigmaC}) was evaluated with this dispersion law
and $\kappa = k_c/2m$ numerically for each system size, and used
then in the definition of the function $\theta (X)$,
Eq.~(\ref{theta}).

We employed the Worm algorithm for classical statistical systems
\cite{Worm} which has Monte Carlo estimators for all quantities of
interest here and does not suffer from critical slowing down. In
cluster methods \cite{Wolff} (which have comparable efficiency)
the calculation of the superfluid density is more elaborate.

From simulations of a series of system sizes and coupling
constants, $U=2,1,0.5,0.25$, we eliminated finite-size and
finite-$U$ corrections to the final results for $\theta (X)$,
$\lambda (X)$, $f_0 (X)$, and $f_s (X)$.

%%%%%%%%%%%%%%%%%%%%%%%%%%%%%%%%%%%%%
\section{Simulation results}
\label{sec:comparison}
%%%%%%%%%%%%%%%%%%%%%%%%%%%%%%%%%%%%%

In Figs.~\ref{figT} and \ref{figN0} we present the final outcome
of simulations throughout the fluctuation region.  We also present
all the data in Table I. The relative accuracy is high far from
the fluctuation region (better then $1~\%$), and gets worse in the
vicinity of the critical point where finite-size corrections are
the largest. We show errorbars in all plots.

\begin{figure}[tbh]
\includegraphics[bb=0 190 650 840 , angle=-90, width=0.4\columnwidth]{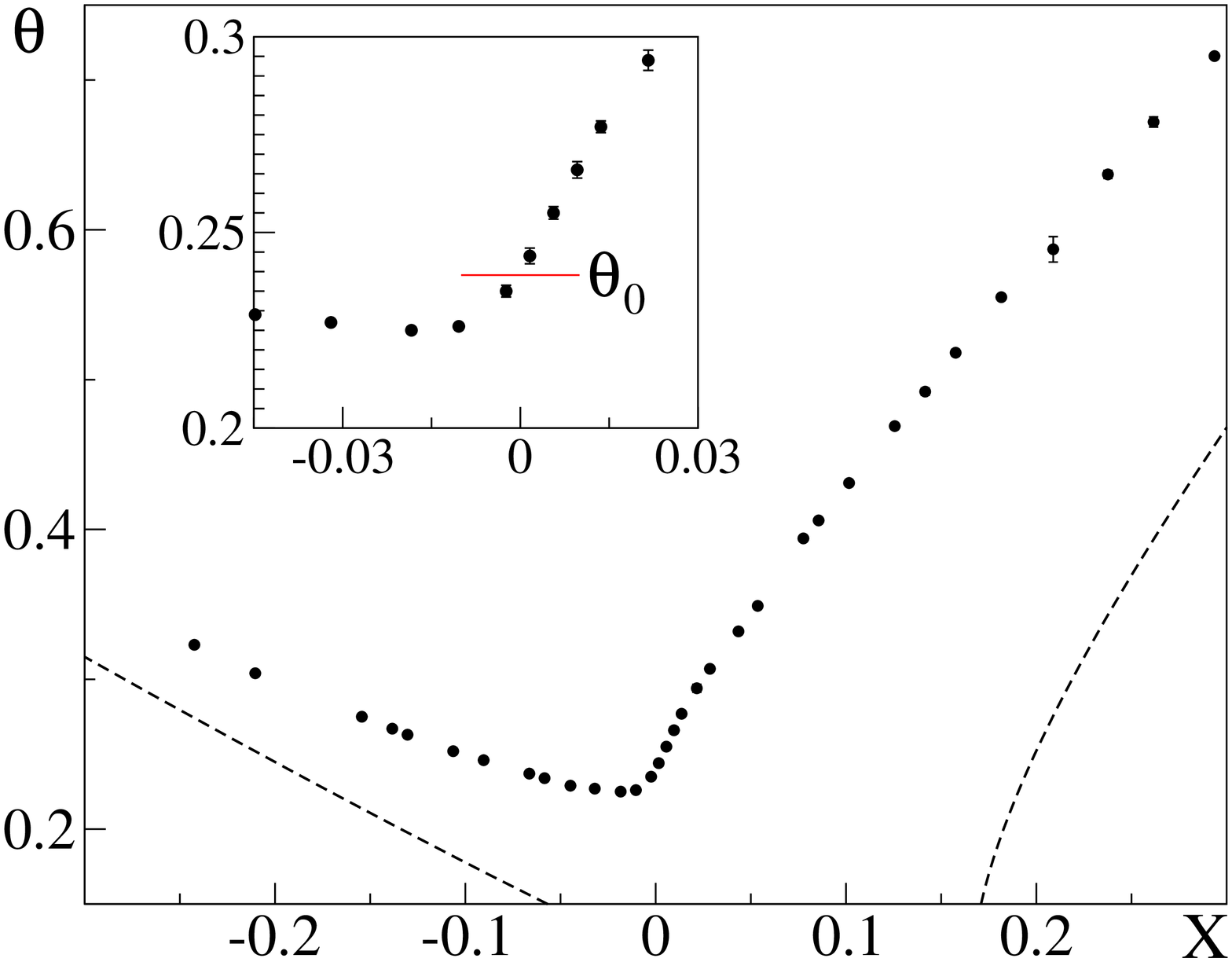}
\includegraphics[bb=0 -0 650 650 , angle=-90, width=0.4\columnwidth]{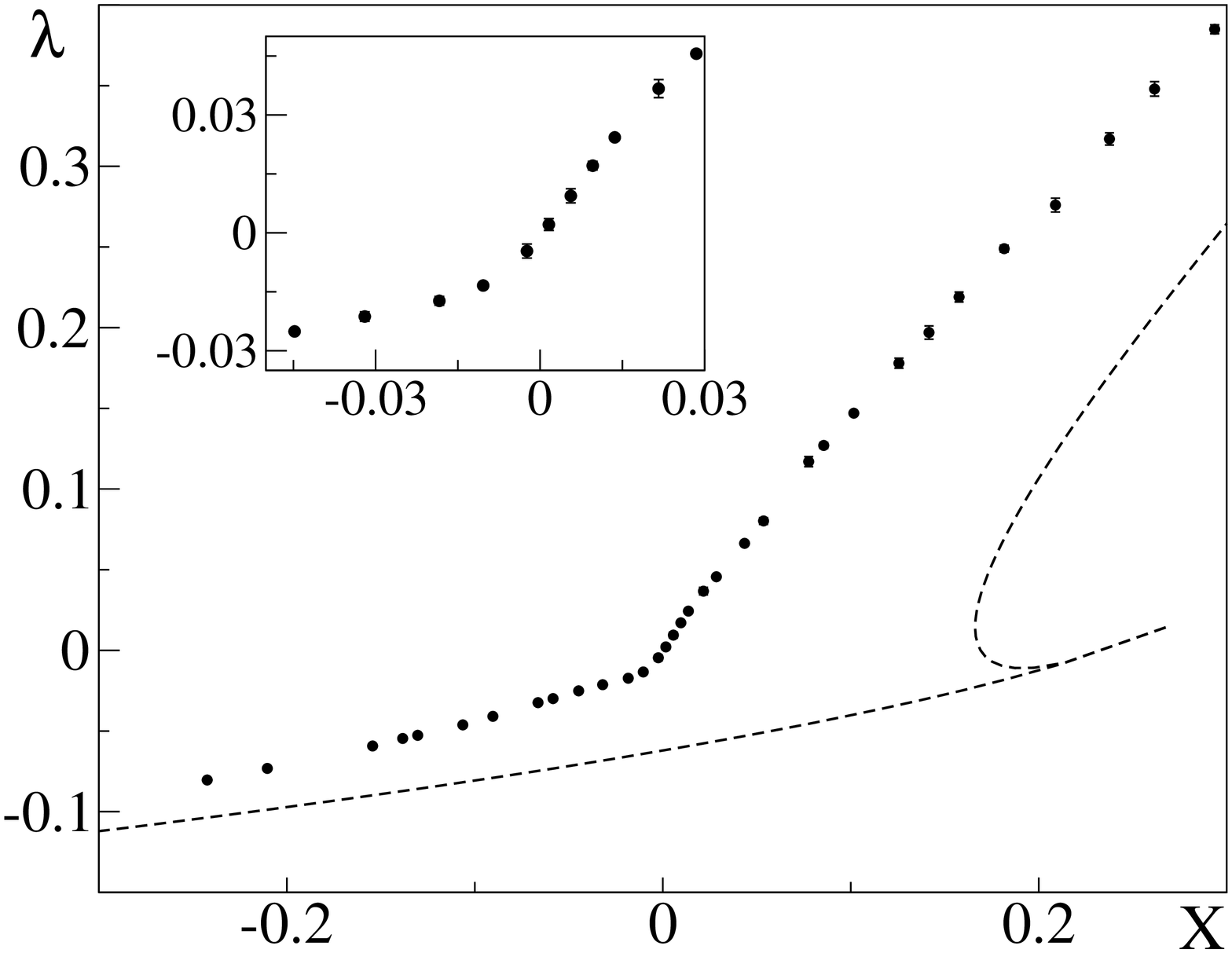}
\caption{The $\theta$- and $\lambda$-functions in the vicinity of
the critical point. } \label{figTinset}
\end{figure}

First, we check for consistency between our approach and the
previously obtained result for $\theta_0$, see Eq.~(\ref{theta0}).
Within the errorbars, the agreement is perfect, see
Fig.~\ref{figTinset}.

Knowing $C$ and $\theta_0$ is all we need to address the
asymptotic MF behavior at large $|X|$, see Eqs.~(\ref{lim0}),
(\ref{lim1}), (\ref{lim2}), and (\ref{lim3}). The agreement with
the MF theory based on the notion of quasicondensate is
remarkable. Despite the fact that in MF we keep consistently only
large terms $\propto X$ and $\propto \sqrt{X}$, for numeric
reasons the constant term also happens to be accurate. The
self-consistent theory of Ref.~\onlinecite{BH} is claimed to go
beyond conventional MF, but it is not known at present whether it
reproduces correctly the $\sqrt{X }$ terms.

The agreement between the data and MF predictions makes it easy to
estimate the size of the fluctuation region (in terms of $X$) to
be about $ \sim 0.4$  on the superfluid side of the transition,
and roughly two times smaller, $\sim 0.2$, on the normal side.

Other quantities of interest are the universal amplitudes for the
superfluid and condensate densities in the critical regime, when
$f_s =A_s X^{\nu }$, and  $f_0 =A_0 X^{\nu (1+\eta ) }$. Here $\nu
=0.6715$ and $\eta =0.038$ are the correlation length and the
correlation function critical exponents of the XY-universality
class in $d=3$, see, e.g., Ref.~\onlinecite{Campostrini}. By
fitting the data for $f_s$ and $f_0$  at small $X$ to the power
laws with known exponents we obtain
\begin{equation}
A_s = 0.86(5) \; ,\label{As}
\end{equation}
\begin{equation}
A_0 = 0.89(5)\; . \label{A0}
\end{equation}

{\bf $\nu $-effect.} It appears in Fig.~(\ref{figT}) as if the
total density has a cusp at the critical point. However, $n$ is
{\it not} singular at $X=0$, and what looks as a cusp in
Fig.~(\ref{figT}) is in fact a relatively sharp crossover slightly
shifted to the normal side of the transition point, see
Fig.~(\ref{figTinset}).

The crossover itself is predicted by MF  equations since $\theta
(X)$ changes its slope from roughly $-X$ to $X$, see
Eqs.~(\ref{lim00}) and (\ref{lim0}). We are not aware of any
special reason why it has to be so sharp and so close to the
transition point---the slope changes at $X\approx -0.01$ and the
crossover region is only about $\Delta X \sim 0.01$ in width. We
call this surprising non-perturbative result the ``$\nu $-effect''
as suggested by the shape of the $\theta$-function.

%%%%%%%%%%%%%%%%%%%%%%%%%%%%%%%%%%%%%%%%%%%%%%%%%%%%%%%%%%%
%%%%%%%%%%%%%%%%%%%%%%%%%%%%%%%%%%%%%%%%%%%%%%%%%%%%%%%%%%%

%%%%%%%%%%%%%%%%%%%%%%%%%%%%%%%%%%%%%
\section{Discussion}
\label{sec:conclusions}
%%%%%%%%%%%%%%%%%%%%%%%%%%%%%%%%%%%%%

To discuss results in the canonical ensemble setup we need to change from
the chemical potential as a control variable to reduced temperature.
First, we note that our results immediately generalize to the case
when the properly rescaled reduced density
\begin{equation}
Y={ 1 \over m^3T^2U } \, {n-n_c(T) \over n_c(T) } \;,
\label{Y}
\end{equation}
is used as a tuning variable.
Indeed,
\begin{eqnarray}
Y &=&\lambda (X) \nonumber \\
n_{\rm cnd}&=&m^3T^2U \, f_0(X)
\label{n0Y}
\end{eqnarray}
is nothing but the parametric dependence of the condensate density
on density at a fixed temperature [and similarly for $f_s(Y)$
and $\theta(Y)$].

%%%%%%%%%%%%%%%%%%%%%%%%%%%%%%%%%%%%%%%%%%%%%%%%%%%%%%%%%%%
\begin{figure}[tbh]
\includegraphics[bb=0   150   650   800, angle=-90, width=0.4\columnwidth]{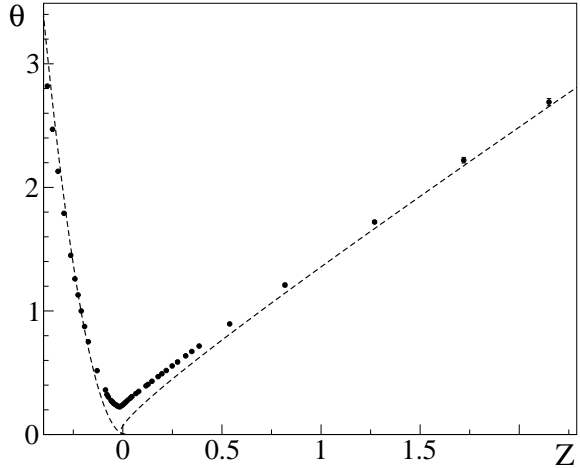}
\caption{The effective chemical potential, $\theta$,
dependence on reduced temperature variable $Z$. }
\label{figTL}
\end{figure}
%%%%%%%%%%%%%%%%%%%%%%%%%%%%%%%%%%%%%%%%%%%%%%%%%%%%%%%%%%%
%%%%%%%%%%%%%%%%%%%%%%%%%%%%%%%%%%%%%%%%%%%%%%%%%%%%%%%%%%%
\begin{figure}[tbh]
\includegraphics[bb=0 190 650 840, angle=-90, width=0.4\columnwidth]{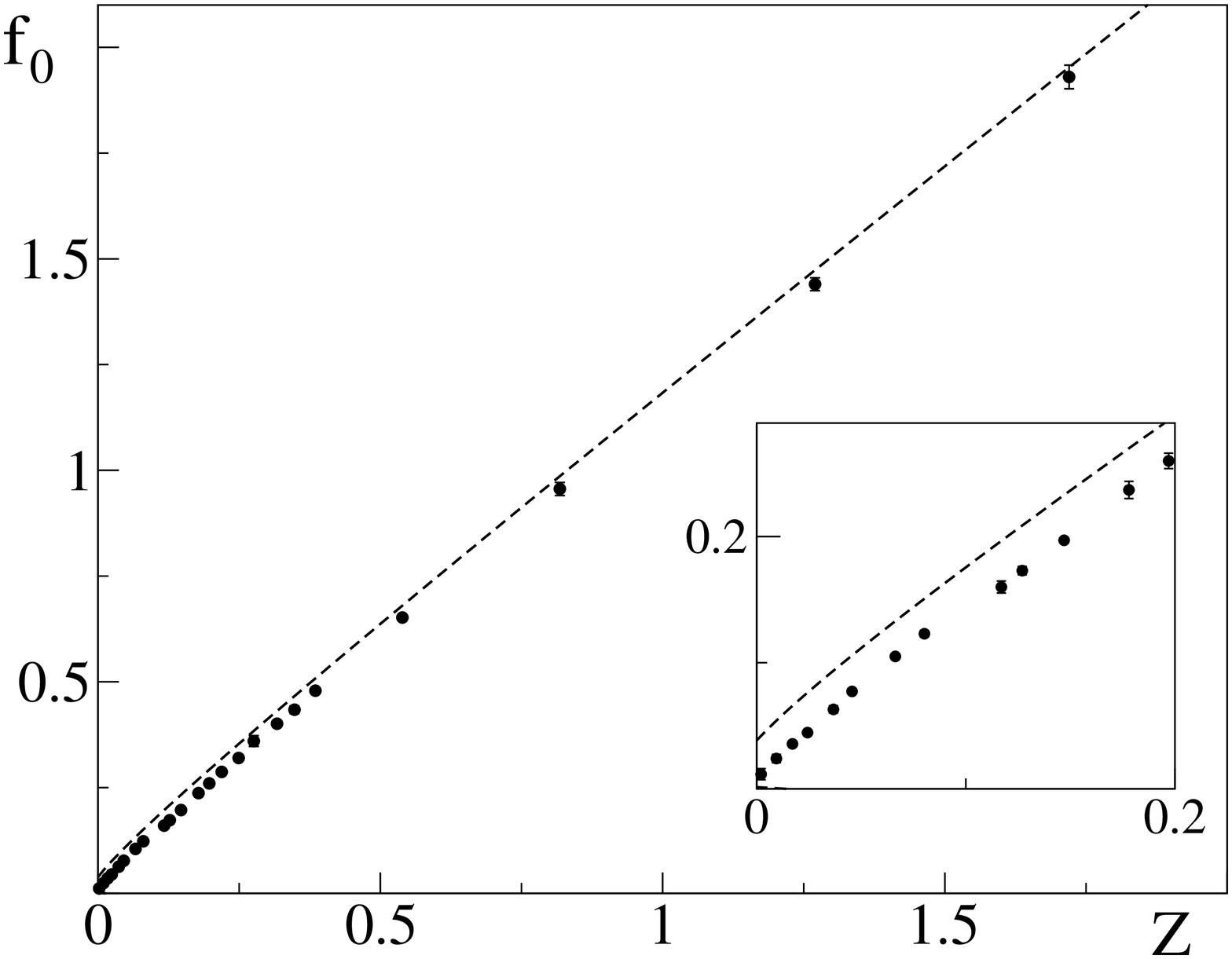}
\includegraphics[bb=0 -0 650 650 , angle=-90, width=0.4\columnwidth]{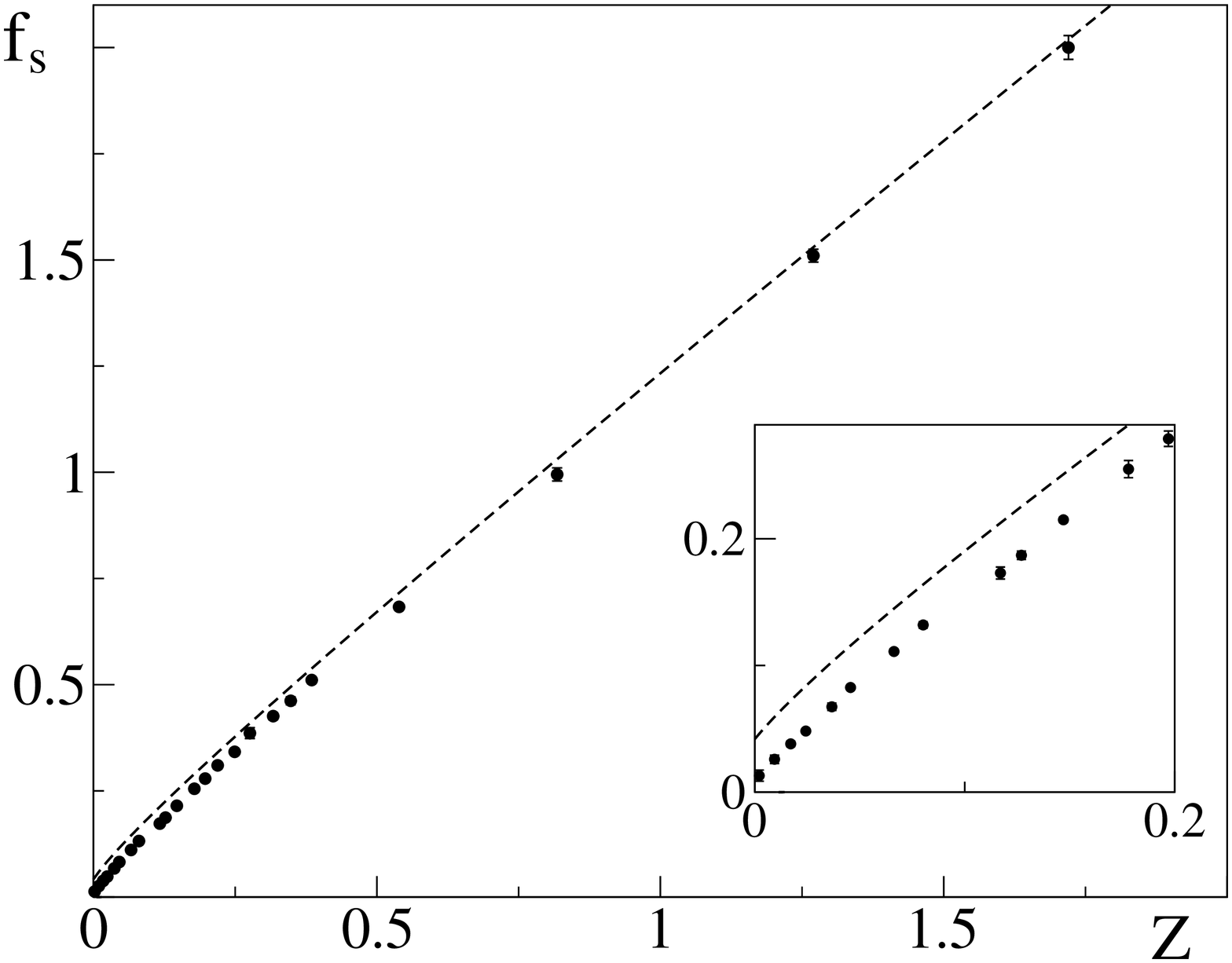}
\caption{The condensate and superfluid density dependencies
on reduced temperature variable $Z$. }
\label{figN0L}
\end{figure}
%%%%%%%%%%%%%%%%%%%%%%%%%%%%%%%%%%%%%%%%%%%%%%%%%%%%%%%%%%%

Next, we observe that Eq.~(\ref{nc}) can be also considered as an
equation for $T_c(n)$
\begin{equation}
\left( {mT_c \over 3.313 } \right)^{3/2} - Cm^3T_c^2U = n  \; .
\label{Tc}
\end{equation}
Using it in the definition of the $\lambda$-function we
obtain
\[
\left( {mT_c \over 3.313 } \right)^{3/2} - Cm^3T_c^2U
-\left( {mT \over 3.313 }  \right)^{3/2}  + Cm^3T^2U =m^3T^2U \, \lambda (X) \; .
\]
Finally, keeping only the leading linear in $U$ and $t$ terms we arrive
at
\begin{equation}
Z = \lambda (X) \; .
\label{ZL}
\end{equation}
where $Z$ is the rescaled reduced temperature parameter
\begin{equation}
Z={3n \over 2 m^3T^2 U}\, t
\; .
\label{Z}
\end{equation}
This establishes the functional relation between $t$ and $X$, and
the parametric dependence of other properties on $t$.

In Figs.~\ref{figTL} and \ref{figN0L} we plot $\theta$, $f_{0} $
and $f_s$ dependencies on the reduced temperature variable $Z$.
From Fig.~\ref{figTL} we deduce the size of the fluctuation region
consistent with the previous estimate in terms of $X$ and relation
$Z=\lambda(X)$, i.e. it is about $\sim 0.5$ on the superfluid side
and $\sim 0.1$ on the normal side. Amazingly, in temperature plots
for $n_{\rm cnd}(T)$ and $n_s(T)$ one may not even distinguish
between the MF theory and numerical data on large scale; only in
the inset which covers just the fluctuation region we can see that
the MF curve goes wrong very close to $T_c$.

We also verify the prediction of the self-consistent theory
\cite{BH} for the critical amplitude of the condensate density. As
mentioned previously, the self-consistent theory makes predictions
for the critical region as well; the relation between the
condensate density and reduced temperature is derived as [see
Eq.~(18) in Ref.~\onlinecite{BH}] $f_0 \approx 0.251\, Z^{\nu '}$
with $\nu ' = 1/( 1+\sqrt{5/6\pi} ) \approx 0.66$. The condensate
density exponent is accurate, though the derivation was done
assuming that the self-energy behaves as $\Sigma(k\to 0) \sim
k^{3/2}$ instead of the true critical behavior $\Sigma(k\to 0)
\sim k^{2-\eta}$ with $\eta=0.0380(4)$ (see
Ref.~\onlinecite{Campostrini} and the discussion below). Since
$d\lambda /dX|_{X=0} = \lambda' \approx 1.8(1)$ the amplitude in
the critical law $f_0 =A_0^{(T)} \, Z^{\nu (1+\eta)}$ is directly
related to the value of $A_0$ in Eq.~(\ref{A0})
\begin{equation}
A_0^{(T)}=A_0 \cdot (\lambda')^{-\nu (1+\eta)} \approx  0.62(4)
\;, \label{A0T}
\end{equation}
in disagreement with the  theory by a factor of two.

To characterize the fluctuation region in terms of the gas
parameter $n^{1/3} a$ we identically rewrite
\begin{equation}
Z= { t \over 91.95\, n^{1/3} a  }
\label{Za}
\end{equation}
The coefficient in the denominator is too large to be ignored even
in qualitative estimates. It was mentioned in Ref.~\onlinecite{us}
that Bose gases should demonstrate universal properties only at $n
a^3 \ll 10^{-5}$. This statement is unambiguously confirmed by the
present study, since for larger values of $n a^3$ the fluctuation
region in temperature is already of order $T_c$ itself. The other
identical way of writing $Z$ is
\begin{equation}
 Z=  { 0.249  \, k_T  \over k_c } t \;.
\label{Zk}
\end{equation}
Obviously, the idea of universality is meaningful only if it is
possible to separate ideal-gas short-wavelength physics from
strongly coupled long-wavelength fluctuations, i.e. when $k_c \ll
k_T$, and thus, according to Eq.~(\ref{Zk}), the fluctuation
region in reduced temperature is small.

In a typical experiment with ultracold gases the parameter $na^3$
is as small as $\sim 10^{-6}$, and the system may be considered as
weakly interacting. The interaction induced critical temperature
shift in the homogeneous system is very small (the corresponding
critical density shift is given by Eq.~(\ref{nc}) and is only
about 1\% for $k_c/k_T \sim 1$) and is difficult to study
experimentally. This does {\it not} imply, however, that all
non-perturbative effects in the vicinity of the critical point are
of academic interest as well.

The quantity directly relevant to the experimental setup is
$\lambda(X)$ since it describes, according to Eq.~(\ref{lambda}),
the density profile in the trapping potential if it is smooth
enough to guarantee the hydrostatic regime. In this regime the
density variation over the mode-coupling radius $r_c \sim 1/k_c$
reduces to $n \equiv n(T, \mu({\bf R}))$, where $n(T,\mu)$ is the
homogeneous equation of state, $\mu({\bf R}) = {\rm const}-V_{\rm
ext} ({\bf R})$, and $V_{\rm ext}$ is the trapping potential. It
follows from Figs.~\ref{figT} and \ref{figTinset} that the
``cusp'' on the density profile does {\it not} coincide with the
point where the condensate first appears, but is slightly shifted
to the perimeter of the trap. Moreover, by fitting experimental
data away from the ``cusp'' using MF equations, one may directly
measure the interaction induced universal chemical potential shift
[through $\theta_0+2C$].

\begin{figure}[tbh]
\includegraphics[bb=0   150   650   800, angle=-90, width=0.4\columnwidth]{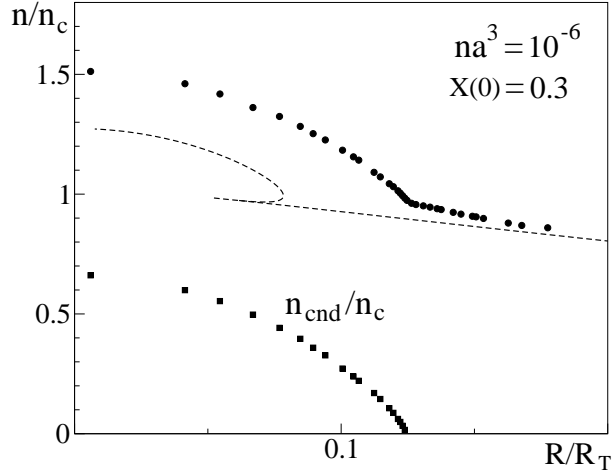}
\caption{The normalized density profile of the weakly-interacting
atomic gas with $na^3=10^{-6}$ in a smooth harmonic trap
(circles). The chemical potential in the middle of the trap
corresponds to $X(R=0)=0.3$ (see text). The dashed line is the
mean-field theory prediction. We also show the normalized
condensate density (squares) for comparison. } \label{trap}
\end{figure}

The variation of the $\lambda (X)$-function across the fluctuation region
is about $\sim 0.3$, which transforms into
density variation of order
\begin{equation}
\Delta n/n_c \sim 40 \, n^{1/3}a  \; ,
\label{experiment}
\end{equation}
or a $40~\%$ strong effect for $na^3 =10^{-6}$!
In Fig.~\ref{trap} we show the density
profile in a smooth parabolic trap
\begin{equation}
V_{\rm ext}\, = \, m\omega^2 R^2 /2 \equiv T (R/R_T)^2 \;,
\label{Vtrap}
\end{equation}
when the condensate density in the middle of the trap is
comparable to the critical density. Amazingly, in this situation
the fluctuation region extends all the way from the critical point
to the trap center, and the MF theory completely fails to describe
the superfluid side. Thus, the non-perturbative physics of the
fluctuation region and the prediction that it is completely
characterized by the classical field theory can be studied even by
experimenting with very dilute gases, $na^3<10^{-6}$. As
Fig.~\ref{trap} clearly demonstrates, all effects in the middle of
the trap are strong.

It is also worth mentioning that in the fluctuation region the
density profile can {\it not} be decomposed into the sum of the
smoothly varying, monotonic non-condensate density background and
the condensate density bump. The normalized condensate density in
Fig.~\ref{trap} increases faster then $n/n_c$, and the naive
``decomposition'' technique would underestimate the actual
condensate density by almost a factor of two.

For the quasi-homogenenous description to work, it is necessary to
keep the external potential gradients small. If the chemical potential
in the middle of the trap corresponds to $(\mu - \mu_c) = m^3T^2U^2 X(0)$,
then the critical point is located at a distance $R_c =( mTU/\omega )\sqrt{2X(0)}$.
The hydrostatic approach can be used when $k_c R_c \gg 1$, or
\begin{equation}
{\omega \over T}  \ll  m^3TU^2 \sqrt{2X(0)} \approx 740 (na^3)^{2/3}\sqrt{X(0)} \;.
\label{omega}
\end{equation}
For the parameters in Fig.~\ref{trap} we need then $ \omega \ll 0.04 T$---
a condition easily satisfied experimentally \cite{gasreview}.
%%%%%%%%%%%%%%%%%% add to references %%%%%%%%%%%%%%%%%%%%%%%%%%%%%%%%
%\bibitem{gasreview} F. Dalfovo, S. Giorgini, L.P. Pitaevskii, and
%S. Stringari, Rev. Mod. Phys. {\bf 71}, 463 (1999).
The crucial point is then in achieving the necessary spatial resolution
in shallow traps.

%%%%%%%%%%%%%%%%%%%%%%%%%%%%%%%%%%%%%%%%%%%%%%%%%%%%%%%%%%%

%%%%%%%%%%%%%%%%%%%%%%%%%%%%%%%%%%%%%%%%%%%%%%%%%%%%%%%%%%%

%%%%%%%%%%%%%%%%%%%%%%%%%%%%%%%%%%%%%%%%%%%%%%%%%%%%%%%%%%%
\begin{figure}[tbh]
\includegraphics[bb=0   150   650   800, angle=-90, width=0.4\columnwidth]{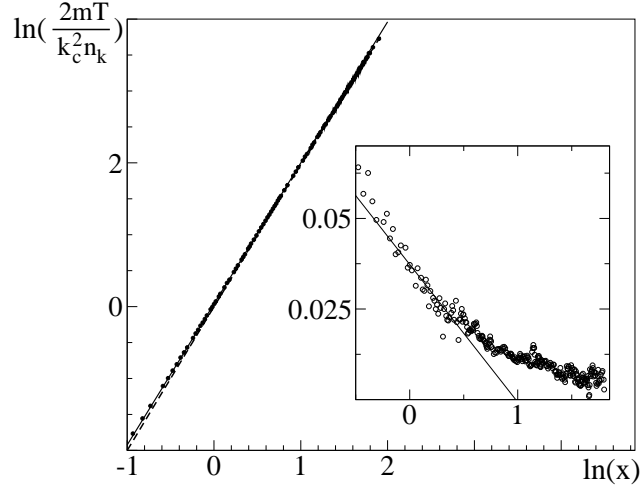}
\caption{The universal part of the occupation number distribution
 $ \ln [ 2mT/k_c^2n_{\bf k}]$ at the critical point.
 Scattering of points is due to statistical errorbars.
The dashed line in the main plot is the bare dispersion law
contribution, $\ln [\epsilon_{\bf k} /(k_c^2/2m)]$, and the solid
line is the critical law $(2-\eta)x+\ln S$. In the inset we plot $
\ln [ T/\epsilon_{\bf k} n_{\bf k}]$ to see the crossover more
clearly. Note  the change in the vertical scale.} \label{figSSS}
\end{figure}
%%%%%%%%%%%%%%%%%%%%%%%%%%%%%%%%%%%%%%%%%%%%%%%%%%%%%%%%%%%

Our final result concerns the universal part of the self-energy at
the critical point \cite{Baym00,BH,ledowski}. The most recent calculation
based on renormalization group equations for the vertex functions
\cite{ledowski} predicted ($x=k/k_c$):
\begin{equation}
\sigma (x) = {2m \over k_c^2}\,
\big[  \Sigma (k_cx) - \Sigma (0) \big]
 \stackrel{k \to 0} \longrightarrow \;
 S  \, x^{2-\eta} \;,
\label{sss}
\end{equation}
with $S=1.54$, $\eta\approx 0.104$, and an extended crossover
between the free particle and critical regimes
(we preserve the present paper definition of  $k_c=m^2TU$).

This quantity is directly related to the occupation numbers in the
$k\to 0$ limit by $(k^2/2mT)\, n_k = 1/[1+\sigma (x)/x^2] $. In
Worm algorithms, the one-particle density matrix is calculated
automatically as the central part of the numerical scheme
\cite{Worm}, and thus $n_k = \sum_{\bf r} \rho ( {\bf r} )
e^{i{\bf k}{\bf r}}$ is readily available. In Fig.~\ref{figSSS} we
plot the universal part of the occupation number distribution as
$\ln [2mT/ (k_c^2 \, n_k ) ]$ versus $\ln (x)$ (in the inset we
plot $\ln [ T/ (\epsilon_{\bf k} \, n_k ) ]$ versus $\ln (x)$,
i.e. subtracting the leading bare spectrum dependence) for the
system size $128^3$ and see the crossover between the free
particle, $1/n_k \propto x^2$, and the U(1) critical, $n_k \propto
x^{2-\eta}$, behavior at $x \sim 1$. The smallest values of $x$
are effected by finite size effects and are not shown in the inset
(in finite systems $n_{k=0}=n_{\rm cnd} \sim 1/L^{1+\eta}$ is
finite at the critical point).

The correlation function exponent is known very accurately \cite{Campostrini}
$\eta \approx 0.0380(4)$, and we
consider it as known (our data are consistent
with this value).
Clearly, very little changes occur
in the structure of the $k^2/2m + \Sigma (k) - \Sigma (0)$ expression
across the fluctuation region. It seems the best way to characterize
the crossover is to write the whole  expression as
$ x^2 + \sigma(x) \equiv x^{2-f(x)}$ with $f(x)$ interpolating between
$0$ and $\eta =0.038$. Also, the crossover is localized in the vicinity
of $x\sim 1$. The numerical value of $S$ in the asymptotic
limit is found to be very close to unity
\begin{equation}
S=1.038(6)
\label{S}
\end{equation}

%%%%%%%%%%%%%%%%%%%%%%%%%%%%%%%%%%%%%%%%%%%%%%%%%%%%%%%%%%%%%%
\section{Acknowledgments}

This work was supported by NASA under grant NAG32870. B.S.
acknowledges a support from the Netherlands Organization for
Scientific Research (NWO).

%%%%%%%%%%%%%%%%%%%%%%%%%%%%%%%%%%%%%%%%%%%%%%%%%%%%%%%%%%%%%%%%
\begin{table}
\caption{\label{tab:table1} Final results for the universal functions
$\theta (X)$, $\lambda(X)$, $f_0(X)$, and $f_s(X)$. }
\begin{ruledtabular}
\begin{tabular}{lllll}
  ~~~~$X  $ ~~~~~~~~~~~~~~~~
& ~~$\theta (X)$~~~~~~~~~~~~
& ~~~~$\lambda(X)$~~~~~~~~~~~~
& ~~~~~$f_0(X)$~~~~~~~~~~~~~
& ~~~~~~$f_s(X)$~~~~~~~~~~  \\ \hline
~    &    ~     &  ~   &  ~ & ~  \\
~-3.738   &    ~3.17     &  ~-0.406(1)  &  ~ & ~  \\
~-3.338   &    ~2.82     &  ~-0.381     &  ~ & ~  \\
~-2.938   &    ~2.47     &  ~-0.355     &  ~ & ~  \\
~-2.538   &    ~2.13     &  ~-0.327     &  ~ & ~  \\
~-2.138   &    ~1.79     &  ~-0.297(1)  &  ~ & ~  \\
~-1.738   &    ~1.45     &  ~-0.263     &  ~ & ~  \\
~-1.498   &    ~1.26     &  ~-0.242     &  ~ & ~  \\
~-1.338   &    ~1.13     &  ~-0.226     &  ~ & ~  \\
~-1.178   &    ~1.00     &  ~-0.210(1)  &  ~ & ~  \\
~-1.018   &    ~0.874    &  ~-0.193     &  ~ & ~  \\
~-0.8584  &    ~0.751    &  ~-0.174     &  ~ & ~  \\
~-0.5384  &    ~0.517    &  ~-0.130(1)  &  ~ & ~  \\
~-0.3064  &    ~0.361    &  ~-0.0875(3) &  ~ & ~  \\
~-0.2424  &    ~0.323    &  ~-0.0804(2) &  ~ & ~  \\
~-0.2104  &    ~0.304    &  ~-0.0732(4) &  ~ & ~  \\
~-0.1544  &    ~0.275    &  ~-0.0593(6) &  ~ & ~  \\
~-0.1384  &    ~0.267    &  ~-0.0546(7) &  ~ & ~  \\
~-0.1304  &    ~0.263    &  ~-0.0527(2) &  ~ & ~  \\
~-0.1064  &    ~0.252    &  ~-0.0462(3) &  ~ & ~  \\
~-0.09039 &    ~0.246    &  ~-0.0409(3) &  ~ & ~  \\
~-0.06639 &    ~0.237    &  ~-0.0324(5) &  ~ & ~  \\
~-0.05839 &    ~0.234    &  ~-0.0299(6) &  ~ & ~  \\
~-0.04479 &    ~0.229    &  ~-0.0251(7) &  ~ & ~  \\
~-0.03199 &    ~0.227    &  ~-0.0213(12)&  ~ & ~  \\
~-0.01839 &    ~0.225    &  ~-0.0173(11)&  ~ & ~  \\
~-0.01039 &    ~0.226    &  ~-0.0134(7) &  ~ & ~  \\
~-0.002393&    ~0.235(1) &  ~-0.0046(18)&  ~ & ~  \\
~~0.001607&    ~0.244(2) &  ~~0.0021(15)&  ~0.0116(30) & ~0.0130(30)\\
~~0.005607&    ~0.255(2) &  ~~0.0094(18)&  ~0.0240(15) & ~0.0259(16)  \\
~~0.009607&    ~0.266(2) &  ~~0.0171(11)&  ~0.0356(10) & ~0.0381(11)  \\
~~0.01361 &    ~0.277(1) &  ~~0.0243(8) &  ~0.0446(7)  & ~0.0482(8)  \\
~~0.02161 &    ~0.294(3) &  ~~0.0367(23)&  ~0.0630(6)  & ~0.0674(7)  \\
~~0.02850 &    ~0.307(2) &  ~~0.0456(8) &  ~0.0772(8)  & ~0.0826(8)  \\
~~0.04350 &    ~0.332(2) &  ~~0.0663(8) &  ~0.105(1)   & ~0.111(1)  \\
~~0.05361 &    ~0.349(1) &  ~~0.0802(21)&  ~0.123      & ~0.132  \\
~~0.07761 &    ~0.394(2) &  ~~0.117(3)  &  ~0.160(2)   & ~0.173(2)  \\
~~0.08561 &    ~0.406(1) &  ~~0.127(2)  &  ~0.173(1)   & ~0.187(2)  \\
~~0.1016  &    ~0.431(1) &  ~~0.147(4)  &  ~0.197(2)   & ~0.215(1)  \\
~~0.1256  &    ~0.469(1) &  ~~0.178(3)  &  ~0.237(4)   & ~0.255(3)  \\
~~0.1416  &    ~0.492(2) &  ~~0.197(4)  &  ~0.260(2)   & ~0.279(3)  \\
~~0.1576  &    ~0.518(2) &  ~~0.219(3)  &  ~0.287(4)   & ~0.310(6)  \\
~~0.1816  &    ~0.555(1) &  ~~0.249(2)  &  ~0.320(3)   & ~0.342(4)  \\
~~0.2089  &    ~0.587(8) &  ~~0.276(4)  &  ~0.360(8)   & ~0.386(5)  \\
~~0.2376  &    ~0.637(3) &  ~~0.317(4)  &  ~0.401(4)   & ~0.426(9)  \\
~~0.2616  &    ~0.672(3) &  ~~0.348(5)  &  ~0.434(5)   & ~0.462(4)  \\
~~0.2936  &    ~0.716(2) &  ~~0.385(3)  &  ~0.479(2)   & ~0.511(6)  \\
~~0.4216  &    ~0.895(2) &  ~~0.539(3)  &  ~0.652(3)   & ~0.683(7)  \\
~~0.6616  &    ~1.21(1)  &  ~~0.818(9)  &  ~0.956(7)   & ~0.995(10) \\
~~1.062   &    ~1.72     &  ~~1.27(1)   &  ~1.44(1)    & ~1.51(1)  \\
~~1.462   &    ~2.22(1)  &  ~~1.72(1)   &  ~1.93(1)    & ~2.00(2)  \\
~~1.862   &    ~2.69(1)  &  ~~2.15(2)   &  ~2.40(1)    & ~2.47(3)  \\
~~2.662   &    ~3.63(2)  &  ~~3.01(3)   &  ~3.28(4)    & ~3.39(7)  \\
~~3.462   &    ~4.56(3)  &  ~~3.89(4)   &  ~4.20(3)    & ~4.33(5)  \\
\end{tabular}
\end{ruledtabular}
\end{table}
%%%%%%%%%%%%%%%%%%%%%%%%%%%%%%%%%%%%%%%%%%%%%%%%%%%%%%%%%%%%%%55

\end{document}